\documentclass{aa}
\usepackage{graphicx}
\usepackage{txfonts}
\begin{document}
   \title{Analysis and interpretation of a fast limb CME with eruptive prominence,
   C-flare and EUV dimming}

   \author{  S. Koutchmy  \inst{1}
 \and
         V. Slemzin \inst{2}
 \and
         B. Filippov \inst{3}
 \and
          J.-C. Noens \inst{4}
 \and
          D. Romeuf \inst{5}
 \and
          L. Golub \inst{6}.  }

   \offprints{S. Koutchmy}

   \institute{ Institut d'Astrophysique de Paris, CNRS and Univ. P. \& M.
   Curie, 98 Bis Boulevard Arago, F-75014 Paris, France \\
              \email{koutchmy@iap.fr}
    \and P.N. Lebedev Physical Institute, Leninsky pr. 53, Moscow,
119991,  Russia \\ \email{slem@mail1.lebedev.ru}
    \and Pushkov Institute of Terrestrial Magnetism, Ionosphere and
   Radio Wave Propagation, Russian Academy of Sciences (IZMIRAN),
    Troitsk Moscow Region, 142190, Russia \\
              \email{bfilip@izmiran.troitsk.ru}
    \and  OMP and Pic du Midi Observatory, France \\
\email{noens@ast.obs-mip.fr}
    \and C.R.I. Claude Bernard Lyon I University, O.A.-Fiducial,
    France \\
    \email{David.Romeuf@recherche.univ-lyon1.fr}
    \and  Harvard-Smithsonian Center for Astrophysics, 60 Garden Street MS58,
    Cambridge, MA 02138 USA \\
    \email{golub@cfa.harvard.edu}
            }

   \date{Received July 00, 2007 ; accepted }


  \abstract
   {}
   {Coronal mass ejections or CMEs are large dynamical solar-corona events.
The mass balance and kinematics of a fast limb CME, including its
prominence progenitor and the associated flare, will be compared
with computed magnetic structures to look for their origin and
effect.}
   {Multi-wavelength ground-based and spaceborne observations are used to study
a fast W-limb CME event of December 2, 2003, taking into account
both on and off disk observations. Its erupting prominence is
measured at high cadence with the Pic du Midi full H$\alpha$
line-flux imaging coronagraph. EUV images from SOHO/EIT and
CORONAS-F/SPIRIT space instruments are processed including
difference imaging. SOHO/LASCO images are used to study the mass
excess and motions. Computed coronal structures from extrapolated
surface magnetic fields are compared to observations.}
   {A fast bright expanding coronal loop is identified in the region recorded
   slightly later by GOES as a C7.2 flare, followed by a brightening and an
   acceleration phase of the erupting material with both cool and hot components.
The total coronal radiative flux dropped by $\sim$ 7\% in the 19.5
nm channel and
   by 4\% in the 17.5 nm channel, revealing a large dimming effect at and above the limb
over a 2 hour interval. The typical 3-part structure observed 1 hour later
by the Lasco C2 and C3 coronagraphs shows a core shaped similarly
   to the eruptive filament/prominence. The total measured mass of the escaping CME
   ($\sim 1.5\cdot10^{16}$~g from C2 LASCO observations) definitely exceeds the estimated
   mass of the escaping cool prominence material although assumptions made to
   analyze the H$\alpha$ erupting prominence, as well as  the
    corresponding EUV darkening of the filament observed several days before, made
    this evaluation uncertain by a factor of 2. This mass budget suggests that the event
 is not confined to the eruption region alone. From the current free extrapolation
we discuss the shape of the magnetic neutral surface and a possible
    scenario leading to an instability, including the small scale dynamics inside and
    around the filament.    }
   {}

\titlerunning{Fast limb CME with eruptive prominence}

   \maketitle
%

\section{Introduction}
Coronal Mass Ejections (CMEs) were originally defined as large
coronal dynamical phenomena (coronal transients) propagating
outward in the field of view of white-light (W-L) externally
occulted space-borne coronagraphs and often showing three parts
(see Wagner \cite{Wagner}). Their primary property is the mass
transport of a presumably coronal mass originally situated behind
the occulting device, toward the external part, without specifying
the origin of this mass revealed by the W-L event due to the
scattering on free electrons. Modelers had to address the
fundamental question: from where does this large amount of coronal
plasma originate? Although many statistical studies of CMEs were
published in recent decades (among the latest, see Vourlidas et
al. \cite{Vourlidas}; Cremades \& Bothmer \cite{ Cremades};
Gopalswamy \cite{Gopalswamy06}) this question has not been
systematically considered and no clear answer exists. One
possibility is that the mass of the CME comes from a related
erupting prominence (EP) and its immediate surroundings. This
assumption was seriously considered recently and prompted another
question: what is the mass of the EP (see Gilbert et al.
\cite{Gilbert05, Gilbert06})? However, the general consensus is
that only the core of the 3-part CME corresponds to the
prominence.

   \begin{figure}
   \centering
    \includegraphics[width=9.9cm]{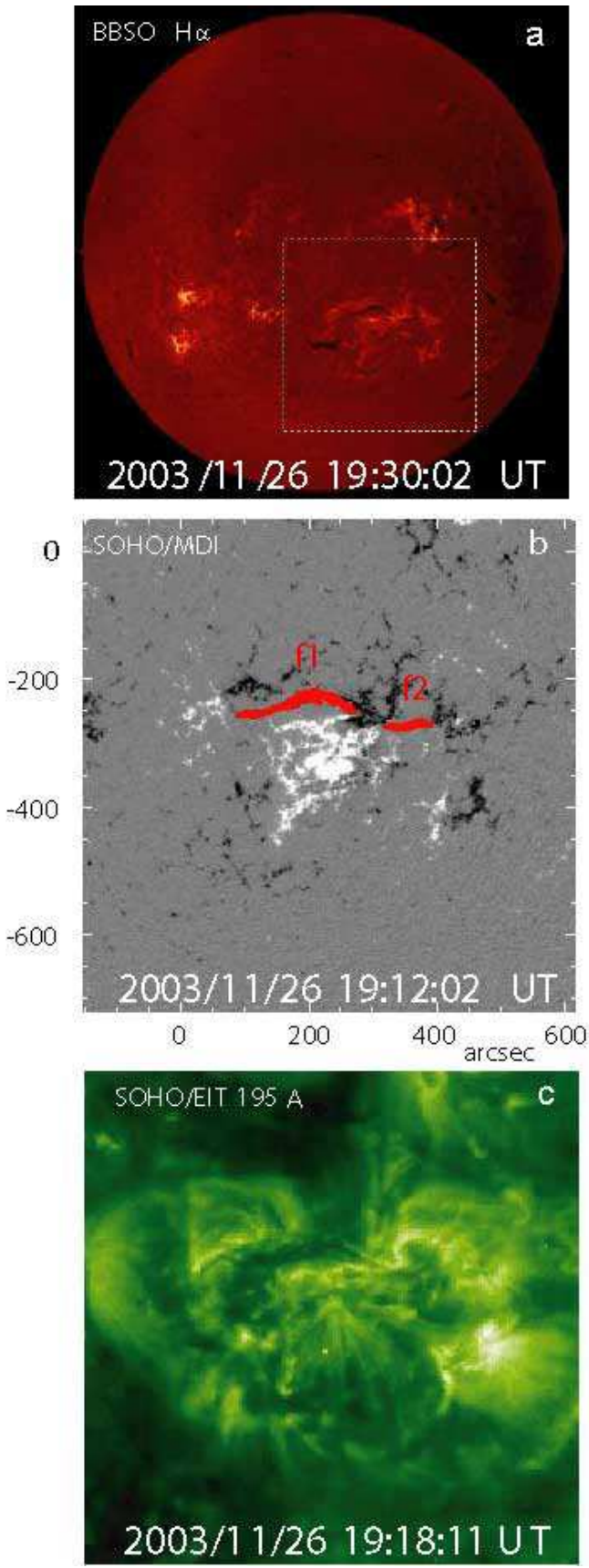}
   \caption { The BBSO H$\alpha$ image (a) showing the zoom box  corresponding to  the MDI magnetic
   map (b) of the AR 10508 region with superimposed contours of filaments (red) from
   the BBSO H$\alpha$ image; (c) EIT image of the box region at
   195 \AA. All images are taken on November 26, 2003.  }
              \label{Fig1}%
    \end{figure}

A distinction has been made in considering the presumably more
important fast CMEs, usually related to a flare and more likely to
be geoeffective and with space weather implications (see e.g. Bao
et al. \cite{Bao}). The EP connection is then sometimes omitted in
favor of a discussion of the origin of solar energetic particles
that contribute only a little to the mass budget.

CMEs are often associated with erupting filaments or prominences
even when considered far from the Sun (see Bothmer \& Schwenn
\cite{Bothmer}). In the last few years many investigations were
undertaken to study the relationship between prominence activity
and CMEs in different wavelengths. Based on H$\alpha$ observations
at the Mauna Loa Solar Observatory, Gilbert et al.
(\cite{Gilbert00}) proposed a classification of active and
eruptive prominences according to their kinematical properties
(radial height, velocity and acceleration). A separation between
escaping prominence material lifting away from the Sun and
material returning toward the solar surface would occur in the
height range from 1.20~$R_\odot$ to 1.35~$R_\odot$ which could be
related to the formation of an X-type neutral line in this region.

Gopalswamy et al. (\cite{Gopalswamy03}) studied the temporal and
spatial relations between the signatures of prominence eruptions
detected in microwaves with the Nobeyama Radioheliograph and
white-light CMEs. It was shown that in most of the studied events
the prominence moved predominantly in the radial direction, with
CMEs and  EPs starting at nearly the same time.  Jing et al.
(\cite{Jing}) presented a statistical study of filament eruptions
observed in H$\alpha$ at the Big Bear Solar Observatory with other
phenomena of solar activity. They found that most eruptions ($>$
50\%) were associated with white-light CMEs, active region
filament eruptions being more associated with flares than
quiescent filament eruptions. Because of their close relationship
to CMEs, EPs were regarded as one of the main sources of material
in the mass balance of CMEs. Indeed, from the theoretical modeling
point of view, the mass of the prominence could be an issue when
considering the launch of the CME (Low et al. \cite{Low}) or, more
generally, its associated magnetic structure (see Chen et al.
\cite{Chen}). Further out, there are strong indications that the
core in the three-part CME structure is associated with a
prominence but this relation is not direct; multi-wavelength
observations are needed.

Evaluations of the EP mass are traditionally based on observations
of the H$\alpha$ absorption of the corresponding filament on the
disk prior to eruption, where they are seen as dark objects. The
destabilization of filaments appears as a sudden disappearance in
H$\alpha$ or {\it a disparition brusque} (DB) (Tandberg-Hanssen
\cite{Tandberg-Hanssen}; Soru-Escaut \& Mouradian
\cite{Soru-Escaut}) which can be caused by an increase of the
plasma ionization degree (thermal effect) or which can also be a
result of Doppler - Fizeau effects (dynamical DB) including a
darkening produced by rapid mass motion (Heinzel \& Rompolt
\cite{Heinzel87}). With TRACE usually a large, high but faint
$\sim$1 MK loop moves outward before anything else happens.

\begin{figure}
   \centering \resizebox{\hsize}{!}{\includegraphics{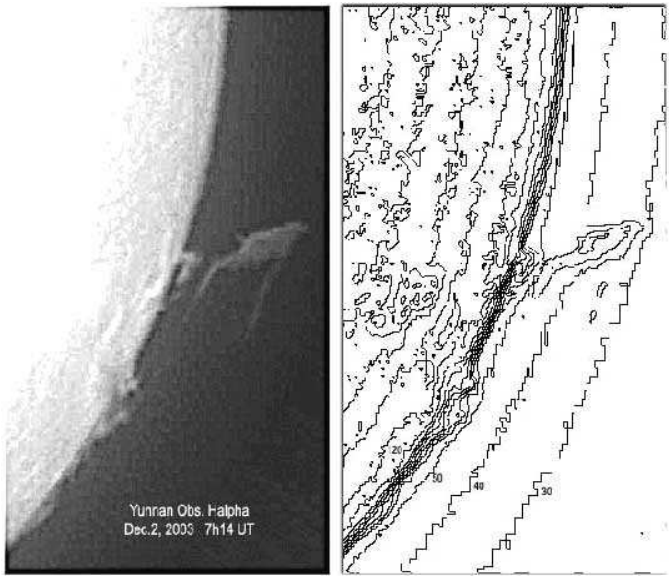}}
   \caption{Partial frame from an H$\alpha$ filtergram taken before the
   eruption from the International H$\alpha$ Network and, at left, the
   deduced set of calibrated isophotes used to evaluate the mass
   of the off-disk prominence. Note the levels of scattered parasitic
   light outside the solar disc (some typical intensities are shown
    near the bottom part).  }
              \label{Fig2}%
    \end{figure}

The approach to estimating the hydrogen density in filaments from
H$\alpha$  absorption was developed by Mein et al. (\cite{Mein96})
and Heinzel et al. (\cite{Heinzel99}) who used the non-LTE cloud
model with optical thickness, Doppler width and velocity as input
parameters. Because the hydrogen ionization degree in the non-LTE
model is a free parameter, the total mass of the filament cannot
be determined unambiguously. When the prominence is seen in
emission outside the disk, a more direct evaluation can be
attempted using both narrow passband W-L coronagraphic
measurements and H$\alpha$  line-flux emission imaging (see
Koutchmy \& Nikolsky \cite{Koutchmy81}). To make filtergrams, a
narrow passband Lyot  filter is usually used and the data need
additional calibrations to take into account the broader line
profile.

This problem was partly overcome by combining H$\alpha$
observations with observations of filaments in coronal EUV lines,
where filaments are seen as dark objects on the background of the
underlying corona, on the disk or even outside the disk. EUV
observations of filaments were started with the Skylab mission and
then continued by SOHO (EIT and CDS) and TRACE. Batchelor \&
Schmahl (\cite{Batchelor}) interpreted the EUV absorption of cool
objects as a continuum absorption of the coronal radiation by
neutral H, and by neutral and ionized He depending on the
wavelength. Kucera et al. (\cite{Kucera}) and Penn (\cite{Penn})
used CDS data for prominence diagnostics and obtained a column
density of neutral hydrogen \ion{H}{i} of $\sim 4\cdot10^{17} -
10^{18}$~cm$^{-2}$. Mein et al. (\cite{Mein01}) have shown that a
combination of H$\alpha$, \ion{Ca}{ii} (8542 \AA) and TRACE 171
\AA\ data may yield a more self-consistent determination of the
neutral hydrogen column density, of the electron density and of
the electron temperature.

\begin{figure*}
\centering
    \includegraphics[width=\textwidth]{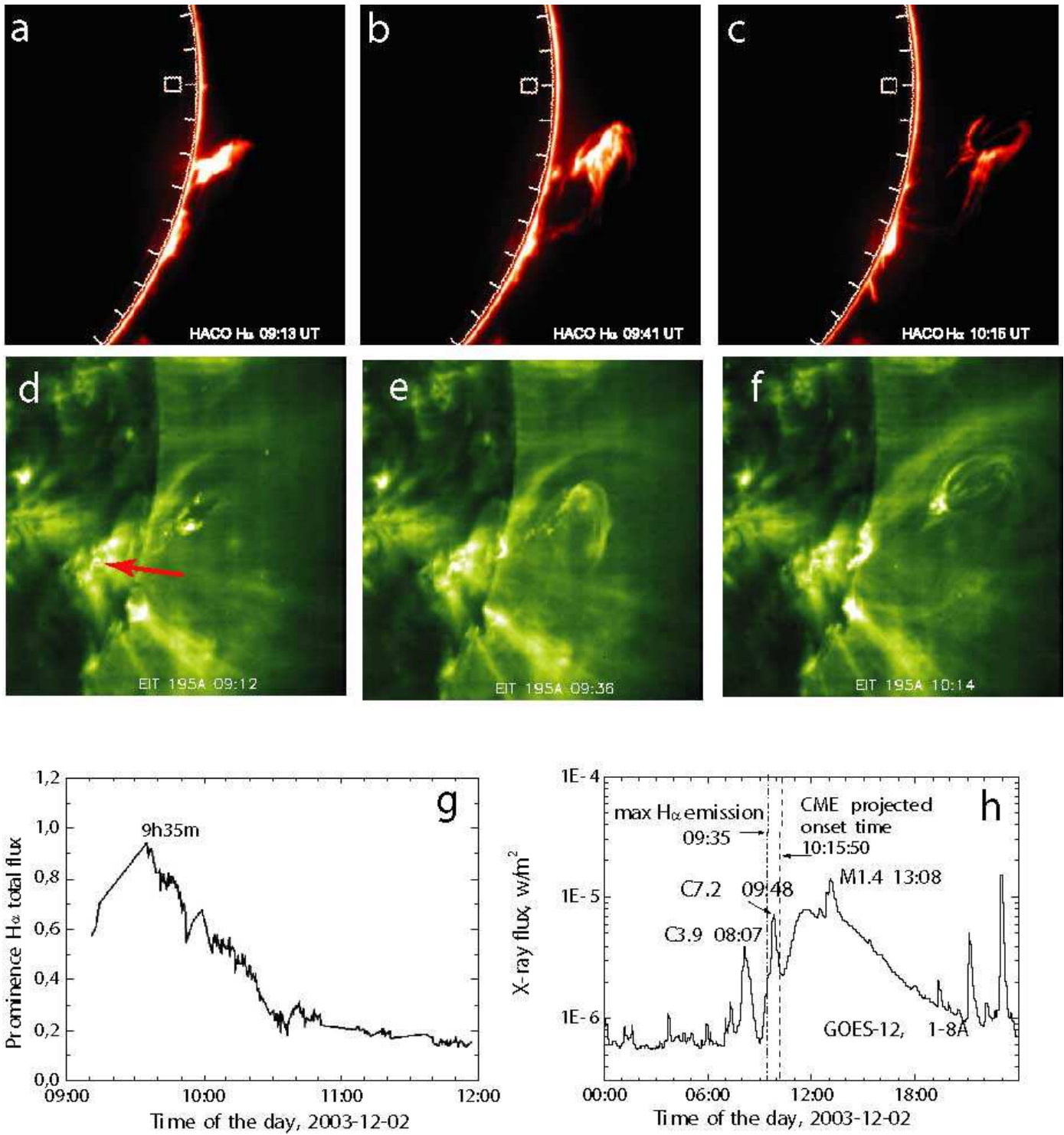}
      \caption{Active phase of the prominence eruption seen with HACO
      in  H$\alpha$ (a)-(c) and in EIT 195 \AA\ (d)-(f). The arrow marks
      the possible place of emergence of the magnetic loop. In the bottom panels,
      the time variations of the total flux in H$\alpha$ measured with HACO (g)
      and, over the full day time interval, in X-rays 1-8 \AA\ measured with
      GOES-12 (h). }
              \label{Fig3}%
    \end{figure*}

Heinzel et al. (\cite{Heinzel01}) and Schmieder et al.
(\cite{Schmieder03, Schmieder04}) noticed that filaments are more
extended in EUV spectral lines with wavelengths below the hydrogen
Lyman-continuum edge (912 \AA) than in H$\alpha$ and they
introduced the term ``EUV filament channel" (EFC) to describe
their observations. A spectroscopic model of the EUV absorption of
filaments and prominences was suggested by Heinzel et al.
(\cite{Heinzel03a, Heinzel03b}),  Anzer \& Heinzel (\cite{Anzer}),
and Schwartz et al. (\cite{Schwartz}). The method takes the
absorption of EUV-line radiation by the Lyman continuum
self-consistently into account, as well as the volume-blocking
effect that is potentially important for coronal lines. A similar
technique  for deriving prominence mass from EUV absorption
features was developed by Gilbert et al. (\cite{Gilbert05,
Gilbert06}). However, when the EUV filament is seen as a
prominence on the off-disk coronal background, often an
edge-brightening around the prominence is observed, strongly
suggesting that heated coronal material is concentrated at the
periphery of the prominence. De Boer et al. (\cite{Boer}) analyzed
the periphery of a prominence using transition region (TR) line
emission and describes this enhanced emission with non-thermal
broadenings and velocities reaching values up to 45 km/s, which is
considerably higher than what is observed using the profiles of
cool lines inside the prominence. During the onset of our EP we
observed a similar coronal edge brightening.

Here we report on a large W-limb event, which occurred on December
2, 2003, 9-14$^{\mbox{\small h}}$ UT, in the vicinity of the old
multi-polar region AR 10508 which was well observed earlier but
was still on the disk at the time of the CME. We analyse the
observations of the prominence eruption taken in H$\alpha$
full-line images with the HACO coronagraph of the Pic du Midi
Observatory (Romeuf  et al. \cite{Romeuf}), images of the
surrounding corona in EUV with SOHO/EIT (Delaboudiniere et al.
\cite{Delaboudiniere}),  and CORONAS-F/SPIRIT telescopes (Zhitnik
et al. \cite{Zhitnik}), and W-L images of the associated CME taken
with the LASCO (SOHO) C2 and C3 coronagraphs. We discuss in detail
the spatial and the temporal development of the eruption, evaluate
the kinematics and especially the mass budget of the CME-related
phenomenon and we propose a possible scenario of the event. This
approach is quite similar to the early work of Plunkett et al.
(\cite{Plunkett}) where a fast limb CME event with an EP was
studied, although no flare was reported (possibly because the
region was already beyond the limb). Here we pay special attention
to the mass budget, including the measurement of the EUV dimming
effect, and we propose a model applicable to this particular event
based on a computed magnetic field context to describe the whole
event, instead of proposing a universal 2.5 D model.


\section{Observations}

\subsection{The context}

A group of three massive filaments was observed one week before
the event by the Global High-Resolution H$\alpha$ Network
observatories (BBSO image in Fig.~\ref{Fig1}a) in the vicinity of
NOAA AR 10508, which is one rotation after the famous AR 10486
that produced extremely powerful X17 flares on October 29 and X28
on November 4 during the preceding Carrington rotation. The
largest filament (which we will  examine further) located to the
north of the center of the AR, was divided into two parts by a
strong negative magnetic field patch seen with MDI
(Fig.~\ref{Fig1}b). The filament was seen in all EUV wavelengths
with EIT and with SPIRIT in the 175~\AA\ and 304~\AA\ bands as a
dark feature. In coronal lines the filament was covered by
overlaying hot loops of the active region. During a week, from
November 25 to December 2, GOES observed more than 20 X-ray flares
in this active region and the filament was not yet destabilized by
the flares.

\begin{table*}
      \caption[]{Geometrical and physical parameters of the filament.}
         \label{T1}
         \begin{tabular}{lllllllllll}
                         \hline \\
            Filament part  &  $D$, cm  &  $S$, cm$^2$  &
             $V$, cm$^3$ & $\tau_{H\alpha}$ & $\tau_{912}$  &  $n_1$,
              cm$^{-3}$ & $n_e$, cm$^{-3}$ & $n_H$,
              cm$^{-3}$ & $M_{H\alpha}$, g  \\
              \\
            \hline
            \\
            f1 & $1.1 \cdot 10^9$ & $2.2 \cdot 10^{19}$ & $2.2 \cdot 10^{28}$ &
            0.48 & 49 & $7.9 \cdot 10^9$ & $1.2 \cdot 10^{10}$ & $2.0 \cdot
            10^{10}$
            & $1.0 \cdot 10^{15}$   \\
            \\
            f2 & $1.1 \cdot 10^9$ & $3.9 \cdot 10^{18}$ & $3.9 \cdot 10^{27}$ &
            0.85 & 96 & $1.5 \cdot 10^{10}$ & $1.6 \cdot 10^{10}$ & $3.1 \cdot
            10^{10}$
            & $2.9 \cdot 10^{14}$   \\
          \\
                     \hline
         \end{tabular}
        \end{table*}

We estimate the geometrical parameters, intensities and derived
physical values of two parts of the filament marked in figure 1b
(top half) as f1 and f2, from its H$\alpha$ image taken on
November 26, 2003 at 17:57:25 UT. The data are shown in
Table~\ref{T1} (the thickness $D$ was not measured and was taken
as equal to the average visible width). Taking into account that
the measured intensity is the sum of partly absorbed background
and foreground, we estimated the optical thickness in H$\alpha$.
Assuming the mean temperature of the filament body (8000~K) and
following the approach of Heinzel et al. (\cite{Heinzel03a}), we
estimate $\tau_{912}$ which gives the neutral hydrogen density
$n_1$, electron (H-ion) density $n_e$ and the total hydrogen
density $n_H$. This gives the mass of both parts of the H$\alpha$
filament $M_{H\alpha} = 1.3\cdot10^{15}$ g. According to Aulanier
\& Schmieder (\cite{Aulanier}), and Heinzel et al.
(\cite{Heinzel03b}), the EFC defined above contains an additional
mass of 50 - 100\% of the H$\alpha$ filament mass, although we are
not sure we see the EFC in our case, see Fig.~\ref{Fig1}. The
total mass of the filament material (H$\alpha$ + EFC) is then
estimated as $M_F = (2.3 \pm .3) \cdot 10^{15}$~g.

\begin{figure}
   \centering \resizebox{\hsize}{!}{\includegraphics{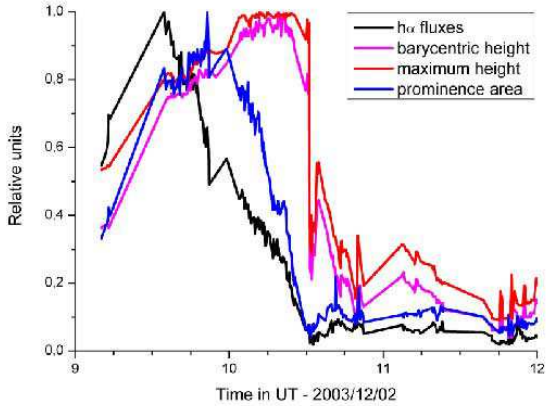}}
      \caption{Temporal variations in relative units of i) the full H$\alpha$
      line flux, ii) the effective surface area of the prominence
      (inside a closed contour drawn at half intensity) and iii)
      the computed heights of the center of gravity and of the
      upper edge of the prominence; after 10:30 UT the values are
      less meaningful because the intensities of the EP drastically drop.
      The phase shifts of the temporal variations of these different
      parameters are illustrated.   }
              \label{Fig4}%
    \end{figure}

\subsection{The erupting prominence and the flare}

On December 1 all day and on December 2 before 09:00~UT, a
filament previously studied on the disk was observed on the west
limb in H$\alpha$  and in EUV lines (with SPIRIT and EIT) as a
quiescent prominence with a center of mass lying at a height of
$\sim$~63 Mm. Later it became an erupting prominence (EP) that we
studied in detail. After 09:10~UT the prominence was observed in
the full H$\alpha$ line with the HACO ({\bf H}-{\bf A}lpha {\bf
CO}ronagraph) imager at Pic du Midi observatory (see Romeuf et al.
\cite{Romeuf} for an exhaustive description of HACO observations)
with a cadence of 15 s. Above the limb a prominence is seen in
H$\alpha$ as a luminous object because of its large opacity to UV
and EUV radiation from the corona, the TR and the chromosphere
(Heinzel \& Rompolt \cite{Heinzel87}). HACO collects the full flux
produced by the H$\alpha$ line (spontaneous transition 3-2 after
level 3 is excited), independently of the dynamic state of the
emitting elements (see Fig.~\ref{Fig3}) thanks to the broad width
(0.25~nm) of the interference filter used by this instrument. This
is in contrast to most instruments taking H$\alpha$ filtergrams
(see Fig.~\ref{Fig2}) which are affected by the Doppler shifts
invariably occurring during the onset of an EP. In addition the
level of scattered parasitic light is greatly reduced in HACO,
compared to using a filtergram (see Fig.~\ref{Fig3}a-c).

At 09:13 UT we observed a new bright small expanding coronal loop
emerging in the center of AR 10508 (indicated by an arrow in
Fig.~\ref{Fig3}d), "striking" the prominence and destabilizing it.
Figure~\ref{Fig3}a-f shows the most important stages of this
process in H$\alpha$ (HACO) and using EIT 195 \AA\ observations.
EUV images synchronously  show the expanding loop and increased
heating of the gas around the prominence body (also called the
periphery, see de Boer et al. \cite{Boer}), which possibly
corresponds to a part of the EUV filament channel. At the same
time the brightness of the prominence body increased in H$\alpha$
to its maximum value (at 09:35 UT, see Fig.~\ref{Fig4}). This is
possibly due to the growth of turbulence inside, which would then
produce a Doppler-brightening effect affecting its small elements
(threads), as Gontikakis et al. (\cite{Gontikakis}) showed using a
detailed calculation for the case of an EP. Following these
authors, the full flux in H$\alpha$ could be magnified by a factor
of 2 to 3, which corresponds to what we observed in the body of
the prominence for a short time, with turbulent velocities which
keep the corresponding shifted line profiles still well within the
pass-band of the filter, provided their transverse components do
not exceed 70 km/s.

\begin{figure*}
   \centering \resizebox{\hsize}{!}{\includegraphics{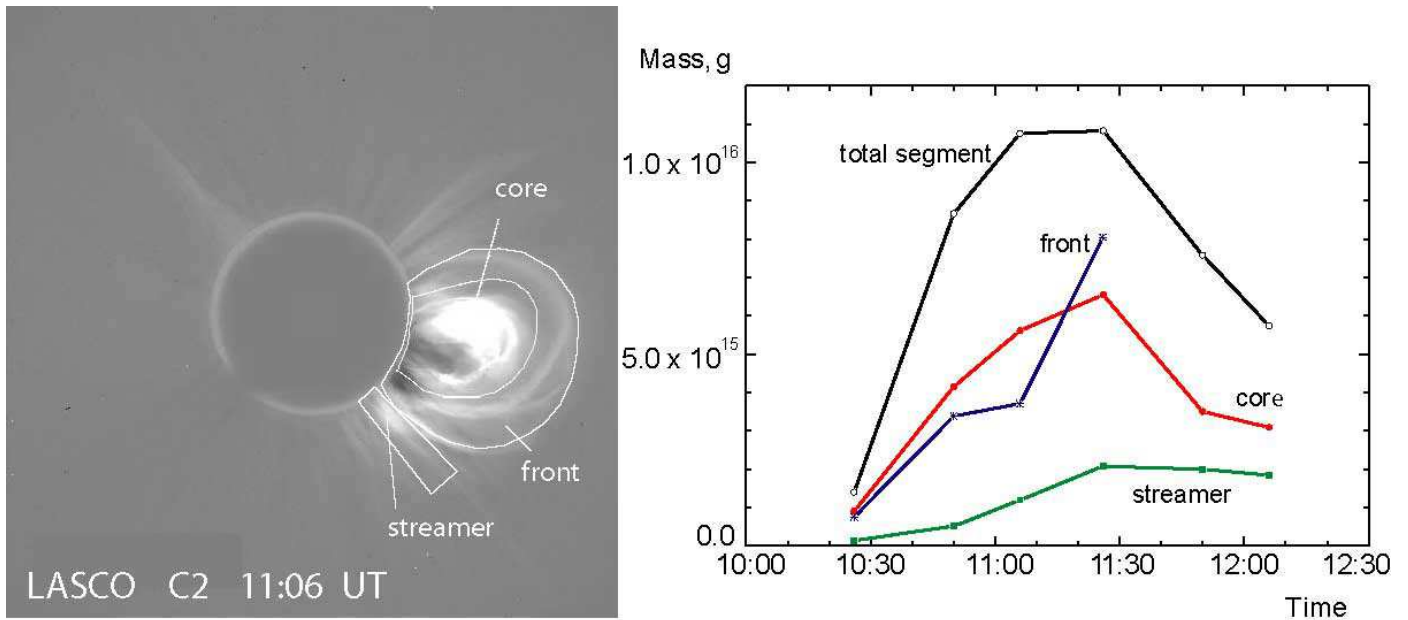}}
      \caption{Estimation of the CME mass from LASCO C2 images: (left) -
      the integrated regions corresponding to the
      frontal structure, the core of the CME and to the streamer;
      (right) - the equivalent mass corresponding to the shown integrated
      intensities. }
              \label{Fig5}%
    \end{figure*}

We looked into the details of these short brightenings using a
simultaneous histogram analysis of each consecutive image taken at
high rate (15 s cadence; not shown here), keeping in mind all
possible variations of instrumental and atmospheric origin
(seeing), to conclude that the prominence was really agitated
``inside". But the details of this analysis are beyond the scope
of this work and we present  here only the summary curves shown in
Fig.~\ref{Fig4} with some small scale rapid variations which could
partly be produced by seeing effects. We also note that the
increase of H$\alpha$ flux is difficult to attribute to a sudden
increase of the density or of the pressure inside the prominence:
neither the morphology nor the kinematics of the EP suggests this.

An increase of the turbulence inside the body of the EP giving an
increase in H$\alpha$ emission, preceding any acceleration upward
(or downward) of the main parts of the EP (see Fig.~\ref{Fig4}),
is a novel feature which tells us something about the heating
processes of the EP before its launch. The measured behavior of
the variations of the effective surface (Fig.~\ref{Fig4}) of the
prominence is an additional argument in favor of the ``turbulent"
scenario because of the observed phase shift between the different
curves: i) variation of fluxes, ii) variations of the effective
surfaces and iii) the motion of the center of gravity of the
prominence, plotted in Fig.~\ref{Fig4}. One possibility to explain
this increase of turbulence inside the prominence is to consider
processes leading to short period (3 to 6 min) oscillations which
would put its threads into rapid transverse motion. Subsequently,
a significant rate of dissipation (thermal heating due to wave
dissipation, for example) would then increase the ionization ratio
and ultimately, will make the EP disappear in H$\alpha$ and make
it then appear in emission in the hotter EUV lines. An argument in
favor of this process comes from the examination of the TRACE
movies of filaments seen in absorption in the 171 \AA\ coronal
channel. They illustrate several cases of agitation phenomena
occurring before the activation (rising) leading to ionization and
eruption (see the filament movies produced by the Lockheed group;
A. Title, 2006, personal communication) of a filament/prominence,
although the limited field of view does not always permit a full
evaluation of dynamical events linked to a CME. In TRACE, bright
emission (\ion{Fe}{ix/x} and \ion{Fe}{xii}) is usually seen
intermixed with the dark absorption during the eruption, not a
long time before (see also Herant et al. \cite{Herant}  Figs. 4, 5
and 6).

\begin{figure}
   \centering
   \includegraphics[width=7cm]{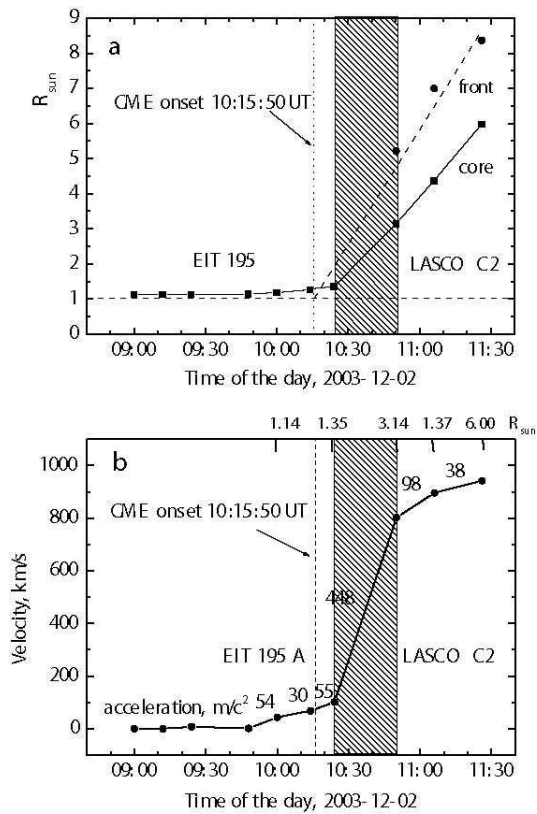}
      \caption{Height-time profile of the prominence, the frontal
      structure and the core of the CME (a); profile of the velocity
      and acceleration of the CME core (b). The shaded area corresponds
      to the time interval where the eruptive material was out of fields
      of view of the instruments. }
              \label{Fig6}%
    \end{figure}

After 09:41 UT the brightness of our prominence decreased in
H$\alpha$ and increased in EUV due to heating and ionization
processes and it started to rapidly move outward, with the hottest
part directed towards the Sun. At 09:48~UT the GOES X-ray
intensity reached its maximum value corresponding to a C7.2 flare.
Figure~\ref{Fig3}e, g shows the light curves of the total flux in
H$\alpha$, measured with HACO and in X-rays (1-8 \AA) measured
with GOES 12.

\subsection{The CME in W-L}

The CME associated with the eruption appeared in the LASCO C2
field of view after 10:50 UT. Before this time (e.g. at 10:26 UT)
a weak streamer was seen nearby at 225$^\circ$. As a result of the
CME development, the streamer at the periphery of the CME was
deflected to the South and its brightness gradually increased with
time. The CME had a typical three-part structure - a frontal
bright loop, a dark cavity and a bright core (see
Fig.~\ref{Fig5}). The shape of the core resembled the shape of the
heated EFC in 195 \AA\ (Fig.~\ref{Fig3}f). After 11:06 UT the
frontal structure and, 20 min later, the core of the CME left the
C2 field of view. In C3 the full CME was seen only in one frame at
12:18 UT.

According to the LASCO catalogue, the CME (frontal structure)
moved in the C2 and C3 from 4 to 25 $R_\odot$ with a nearly
constant velocity of 1393 km/s. At 6$R_\odot$ the core moved with
a velocity of 942 km/s. The points corresponding to the core in
the height-time graph match well the same line as the last point
for the prominence, which suggests a close relationship. The
activation and the motion of the prominence started at least half
an hour before the CME onset determined from a linear
interpolation of the LASCO points.

\section{Mass loading and kinematics of the eruption and the CME}

The first component to consider is the mass directly injected or
loaded by the EP and its surroundings. We discussed in Sect. 2.1
the determination of the mass of the H$\alpha$ filament which was
observed well before in the active region underlying the event
(Fig.~\ref{Fig1}) and evaluated its mass, including its immediate
surrounding, as $(2.3 \pm .3)\cdot10^{15}$ g (this value possibly
may be doubled due to turbulence).

We used several methods to evaluate the more relevant bulk mass of
the prominence becoming the EP just prior to its launch
(Fig.~\ref{Fig2} and Fig.~\ref{Fig3}a). The effective surface area
of the prominence is easy to measure. Then, we compared its
average intensity and morphology with those of typical prominences
that were measured with the aim of determining typical masses of
prominences with the large 52~cm aperture coronagraph of the high
altitude station near Kislovodsk. Narrow pass-band filters (FWHM
of .5~nm) in W-L (a continuum window near 607.3 nm) and around
H$\alpha$ were used to perform a photometric analysis (Koutchmy \&
Nikolsky \cite{Koutchmy81}). We found a typical electron density
integrated along the l.o.s.  (or ``column" density) of the order
of $2.5\cdot10^{38}$ cm$^{-2}$. Assuming an ionisation rate of
50\% over the $10^3$ Mm$^2$  effective surface we have a mass of
$1.2\cdot10^{15}$ g. Inside the prominence the ionisation ratio
could be lower and  the estimated mass of the prominence will
increase by a factor of 2 to 4.

The second method we used takes advantage of intensity
measurements done on the filtergram shown in Fig.~\ref{Fig2}, by
comparing the isophote intensity levels over the prominence image
with the simultaneously observed disk intensities. We then can
obtain the integrated l.o.s. amount of plasma with neutral
hydrogen in level 2, and then assumptions have to be made
concerning the thermal equilibrium among the levels to determine
the abundance of \ion{H}{i} at ground level 1, where densities are
several orders of magnitude higher, etc. At local thermo-dynamical
equilibrium (LTE), it will depend on the excitation temperature
(Gouttebroze et al. \cite{Gouttebroze}) which we take to be of the
order of 6000 K. The deduced column density from the averaged
values of H$\alpha$ fluxes of Fig.~\ref{Fig2} gives an overall
mass, taking into account the extension shown by isophotes, quite
close to the mass deduced from W-L coronagraphic observations,
when a 50\% ionisation ratio is assumed (see also Koutchmy \&
Nikolsky \cite{Koutchmy81}). It could mean that this assumption is
not too far from reality. However, the ionisation ratio is a
matter of discussion: it is certainly not uniform inside the
prominence (and even across the section of a single thread) and at
least at the periphery  where H$\alpha$ emission disappears, the
ratio is close to 100\%. Further, the assumption of LTE is
certainly not satisfied, and finally the temperature is not
uniform (see Wiehr et al. \cite{Wiehr}). In this respect, what the
EP region looks like using  different EIT channels is important.
The 304 \AA\ image of He+, taken several hours before the
eruption, shows the prominence in absorption but surrounded by
rather strong emission; it is similar but weaker in the 171 \AA\
(\ion{Fe}{ix/x}) channel. In the 284 \AA\ channel where the
coronal temperature is much higher (\ion{Fe}{xv}), it is not like
this: although the prominence is also seen in absorption, no
enhancement is seen at the periphery and the coronal emission is
more extended. This tells us that the extended surrounding of the
EP is an important part of the whole eruption and that the mass of
a large part of this area has to be taken into account (see the
discussion of the dimming effect below).

\begin{figure*}
  \centering
    \includegraphics[width=\textwidth]{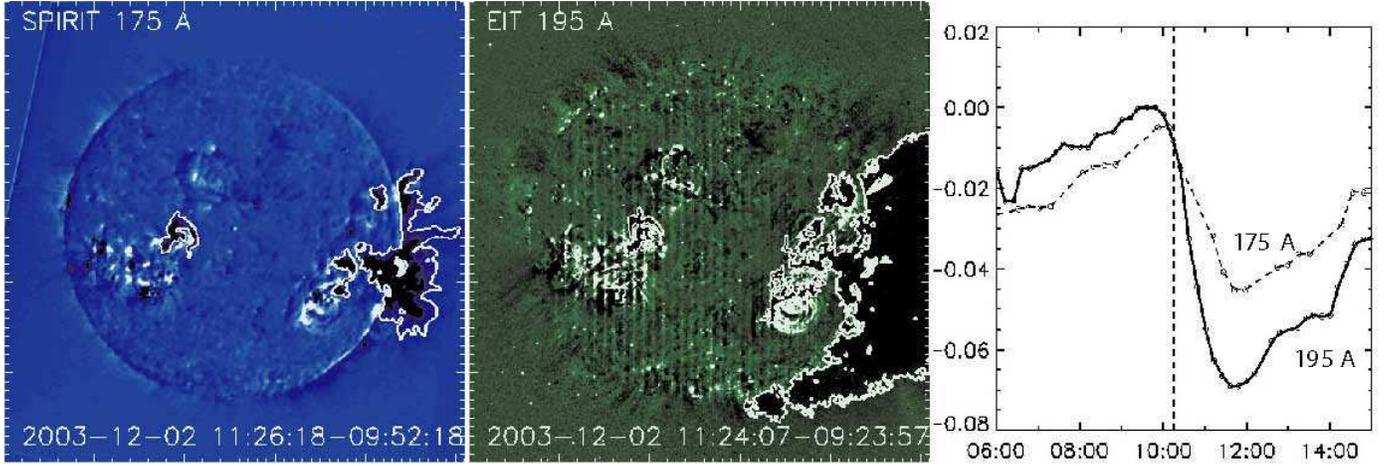}
      \caption{Left and center panels - EUV dimmings in the
      SPIRIT 175 \AA\ and EIT 195 \AA\ difference images. Right panel
      - the total intensity in the dimming area relative to
      the total intensity of the nominal images. Dashed line
      corresponds to the projected LASCO CME onset (10:15:50 UT). }
              \label{Fig7}%
    \end{figure*}

In view of the large dispersion found when trying to determine the
mass of the prominence from the observed H$\alpha$ flux (we do not
discuss errors produced by using averaged values without going
into the details of the optical thickness variations and
variations of the profile of the line), we refrain from giving
formulae and numerical results, aside from the rough estimates
given above. Our conclusion is that more coronagraphic
observations of prominences are needed, such as observing
simultaneously the H$\alpha$ flux and the K \ion{Ca}{ii} flux to
deduce a parameter sensitive to the ionisation ratio of
\ion{H}{i}. The method promoted in EUV by Gilbert et al.
(\cite{Gilbert05, Gilbert06}) should also be developed (up to now,
their sample of prominences gave masses not exceeding $10^{15}$
g).

We also evaluated the mass of the CME using the LASCO images and
the standard SolarSoft procedures, which include the allocation of
regions of interest over which intensities are integrated. In each
image we separately treated three regions, which correspond to the
frontal structure, the core and the streamer (Fig.~\ref{Fig5}a).
The plots of the calculated masses are presented in
Fig.~\ref{Fig5}b.

The estimated values of masses corresponding to the time of the
full CME expansion in the C2 field of view (11:26~UT) are: for the
frontal structure    $8\cdot10^{15}$ g, for the core
$6.6\cdot10^{15}$ g, for the streamer   $2\cdot10^{15}$ g; for the
total amount including all CME parts and the streamer
$1.1\cdot10^{16}$ g. The values for the frontal structure and for
the core partly include the moving streamer component because it
is difficult to clearly distinguish between them. A more realistic
estimate is obtained after subtraction of the streamer mass from
each component, which gives a mass of the frontal structure of
$6\cdot10^{15}$ g, for the core $4.6\cdot10^{15}$ g. The sum of
these values gives a full CME mass for C2 (without the streamer)
of $10.6\cdot10^{15}$ g. The total CME mass value for C3 at 13:42
UT ($R = 25R_\odot$) is estimated as $1.5\cdot10^{16}$ g, showing
that the process of CME development effectively continued.

The total eruption/CME process can be roughly divided into three
stages: activation, eruption of the prominence, and propagation of
the CME in the LASCO field of view. Between the last moment where
the prominence was seen in EUV or H$\alpha$ (10:24 UT) and the
first appearance of the CME in the LASCO C2 field of view (10:50
UT) there is a 'grey zone' where the position of the erupting
matter and the parameters of its evolution are not possible to
determine.

We measured the positions of the moving prominence seen in
absorption from the images taken with the EIT and the SPIRIT
telescopes (from 09:00 UT to 10:24 UT) and combine it with the
measured positions of the CME core in the LASCO C2 images. The
height-time profiles of the prominence/core and frontal structure
are shown in Fig.~\ref{Fig6}a, the derived values of velocity and
acceleration of the core - in Fig.~\ref{Fig6}b. The dotted lines
in these figures correspond to the projected CME onset time from
the LASCO CME catalogue (http://cdaw.gsfc.nasa.gov/CME\_ list)
where it was derived from the height-time profile for the frontal
structure using a linear approximation.

\section{EUV dimmings}

In general, dimmings (areas of temporarily reduced  emission
measure) in EUV or soft X-ray images appear at the places of the
coronal plasma outflow during an eruption leading to a CME (e.g.
Zhukov \& Auchere \cite{Zhukov}). It is interpreted as a density
depletion associated with the CME initiation (Hudson \& Webb
\cite{Hudson}; Zarro et al. \cite{Zarro} and others). The
procedure of image pre-processing and of selection of the dimming
regions we used here was described in detail elsewhere (Chertok et
al. \cite{Chertok04}; Chertok \& Grechnev \cite{Chertok05};
Slemzin et al. \cite{Slemzin05, Slemzin06}). A detailed analysis
of dimmings and their relationship with CMEs have been carried out
in many papers (e.g. Harra \& Sterling \cite{Harra}; Harrison et
al. \cite{Harrison}; Grechnev et al. \cite{Grechnev} and so on).
Maps of dimming regions we obtained in 195 \AA\ and 175 \AA\ are
shown in Fig.~\ref{Fig7}a and Fig.~\ref{Fig7}b. Graphs in
Fig.~\ref{Fig7}c display the light curves of the total dimming
intensities relative to the total intensities in the original (not
differenced) nominal images. The light curves for both cases are
similar: they reached a maximum at $\sim$~09:30~UT - 10:00~UT
before the CME onset, then decreased to a minimum at 11:30~UT -
12:00~UT. The drop of intensity in the dimming area relative to
the total solar intensity before the eruption is large: in 195
\AA\ it is about 7\%, in 175 \AA\ it is about 4.5\%. The spectral
response functions of both bands are quite similar except that a
small, pronounced hot component (10-15~MK) exists in the 195 \AA\
case.

We note that the beginning of the gradual dimming intensity
decrease in 195 \AA\ (Fig.~\ref{Fig7}c) corresponds to the moment
of the launch of the prominence  (dash-dot line in
Fig.~\ref{Fig7}c) and the maximum of intensity of the X-ray flare.
If we assume that the EUV dimming is associated with the formation
of the frontal structure, one can conclude that both the frontal
structure and the core of the CME were initiated very close in
time (within 5-10 min).

\section{Interpretation: the magnetic field structure and the filament equilibrium}

There is no doubt that the filament destabilization and its
eruption were related to the properties and the changes of the
magnetic field in the active region. We first note that the CME
morphology (see Fig. 5) seems to fit one of those identified by
Cremades \& Bothmer (\cite{Cremades}), where the observed
morphology seems to be consistent with what is expected for the
neutral line location and the position of the eruption near the
limb. However, the coronal magnetic field is not measured nor even
qualitatively observed. Accordingly, this part of the work is
mainly based on extrapolations made from the surface of the Sun
and on guesses of the role of the magnetic field. The position of
the active region, close to the limb on December 2, was favorable
for measuring the plasma density distribution and the velocity of
coronal structures but made it almost impossible to see the
changes in the photospheric magnetic field. In order to guess the
magnetic configuration in the erupting region we started from the
time when the region was not far from the center of the disc on
November 25.

\begin{figure}
   \centering
   \resizebox{\hsize}{!}{\includegraphics{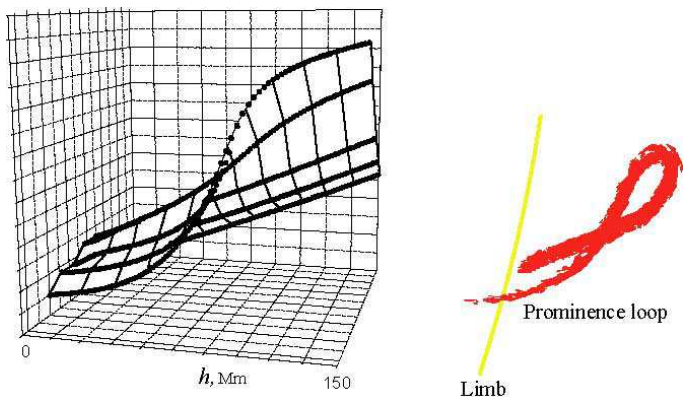}}
   \caption{Projection of the magnetic neutral surface in
   the AR 10508 on November 25, as it should look near the
   limb, calculated above the neutral line related to the
   eruptive prominence (left), and the  side view of a loop
   lying in the neutral surface that models the eruptive
   prominence (right). }
              \label{Fig8}%
    \end{figure}

We calculated the current-free coronal magnetic
field\footnote{Generally speaking, the presence of a non-potential
flux rope is a necessary part of the model. A quite similar
approach was recently attempted in the context of an active region
(van Ballegooijen \& Mackay \cite{van Ballegooijen}).} $\vec B$ in
the vicinity of the filament using the photospheric magnetic field
measurements from SOHO/MDI. As the region of interest was not
large compared with the whole surface of the hemisphere, we
considered the boundary as a flat surface restricted by the size
of the active region containing the main magnetic concentrations.
We used the well-known solution for half-space with a plane
boundary :

\begin{equation}
{\vec B} = \frac{1}{2 \pi} \int \int \limits_S \frac{B_n (x', y',
0)\: {\vec r}}{r^3} \; dx' dy'  ,
\end{equation}
where $B_n$  is the normal magnetic-field component on the plane
$S$ and $\vec r$ is the radius vector from some point on the
surface to a given point in the corona. Zagnetko et al.
(\cite{Zagnetko}) showed that the prominence material is mainly
concentrated at the surface that passes through the apex of the
field line arches which is a magnetic neutral surface. Recall that
the systematic inclination of the bodies of large quiescent
prominences at moderate latitudes toward the west, which had been
widely described by d'Azambuja \& d'Azambuja (\cite{d'Azambuja}),
was explained by the peculiarities of the structure of the
large-scale magnetic fields. We calculated the coordinates of the
neutral surface in AR 10508. Figure 8 shows the projection of the
magnetic neutral surface, as it should look near the limb,
calculated above the neutral line related to the studied filament
up to the height of 150 Mm. We can compare the shape of a nearly
semicircular loop lying in the neutral surface (Fig.~\ref{Fig8},
right) with the shape of the prominence at the early stage of the
eruption (Fig.~\ref{Fig3}b-c) when the loop of the eruptive
prominence roughly conserves its initial form. At equilibrium, the
prominence loop should be located near the neutral surface, hence
the front leg of the prominence should be curved and in the middle
part nearly parallel to the solar surface, while the back leg
should be more straight and tilted from the vertical direction at
an angle of about 30$^\circ$. This is qualitatively the same as
what we see in Fig.~\ref{Fig3}b-c.

\begin{figure}
   \centering
   \resizebox{\hsize}{!}{\includegraphics{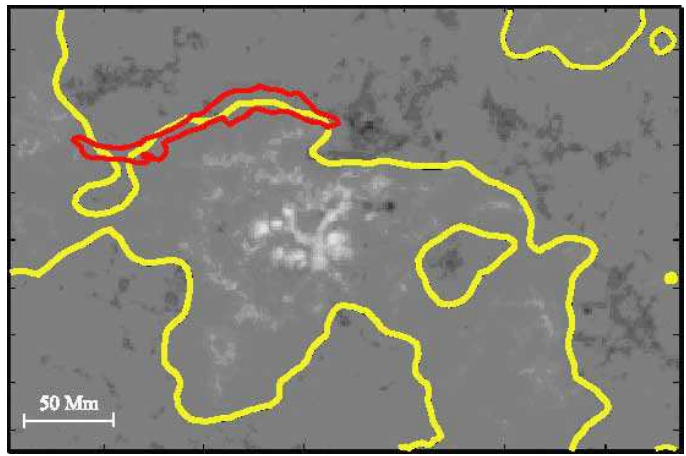}}
   \caption{The neutral line calculated at a level of 6 Mm above
   the photosphere (yellow contours) overlaid on the photospheric
   MDI magnetogram obtained on 25 November 2003 at 19:12 UT within
   the area used as the boundary condition for coronal magnetic
   field calculations. The filament position taken from the BBSO
   H$\alpha$ filtergram for 16:45 UT 25 November 2003 is shown
   as a red contour.}
              \label{Fig9}%
   \end{figure}

Then we found the limiting height of the filament equilibrium
$h_c$ by solving the equation (Filippov \& Den \cite{Filippov00,
Filippov01})

\begin{equation}
\left. \frac{dB}{dh} \right|_{h_c} = - \frac{B(h_c)}{h_c} .
\end{equation}
This parameter is derived from the vertical stability condition in
the inverse polarity filament model (Van Tend \& Kuperus \cite{Van
Tend}; Molodensky \& Filippov \cite{Molodensky}; Forbes \&
Isenberg \cite{Forbes}) and reflects the scale height of the
magnetic field of photospheric sources. We calculated the limiting
height using the SOHO/MDI magnetograms of November 25, 27, and 29.
The results for the two former days are very similar.
Figure~\ref{Fig9} shows the calculated neutral line at the height
of 6 Mm, where the field is smoothed enough, overlaid on the
photospheric magnetogram. The filament that erupts on December 2
is also shown as a white contour. In the middle part of the
filament length, where its axis is rather straight, the limiting
height is higher than the vertical size of our calculation domain
(150 Mm). Near the filament ends, the curvature of the neutral
line increases. This means that the local scale of the
photospheric magnetic field is smaller there. In accordance with
the reduced field scale, the limiting height is lower and reaches
$\sim~60$~Mm. However, near the filament ends field lines are
anchored in the photosphere and the magnetic tension plays an
important role, whereas Equation 2 does not take into account the
anchoring of the filament legs.

On November 29 the AR 10508 was close to the limb and the magnetic
field measurements became unreliable. We then used a special
procedure to obtain the component normal to the solar surface from
the line-of-sight measurements (Den \cite{Den}). Calculations show
a small lowering of the limiting height at the filament ends and a
significant lowering in the middle of the filament down to 40~Mm.
This value possibly reflects the tendency of a lowering of the
limiting height; however, as has been mentioned, the initial
photospheric magnetic field measurements in this region are not
reliable.

Observations of the filament (and the corresponding prominence)
near the limb on November 30 and December 1 demonstrate that the
filament was stable up to a rather great height. On November 30
the height of the filament top can be estimated as 35 Mm, while on
December 1 - as 50 Mm. The greatest height on December 2, before
the filament begins to ascend, was 120 Mm. This value is smaller
than the limiting height on 25-27 November (more than 150~Mm) but
greater than the limiting height on November 29 (40 Mm, possibly
underestimated). It is not unlikely that the height of the
filament reaches a limiting value not long before the eruption, as
has been shown for several tens of filaments by Filippov \&
Zagnetko (\cite{Filippov07}). We believe that the rising of the
filament, possibly due to the growth of an electric current (that
we do not discuss here), within a flux rope containing the
filament, up to the threshold of stability, is the main cause of
the eruption. On the other hand, changes in the active region
magnetic field can lower the threshold. When the height of the
filament reaches the limiting height any small disturbance is able
to initiate the beginning of the eruption.

The sequence of events observed on the west limb on December 2
between 07:30 UT and 11 UT can then be interpreted in the
framework of a model of an erupting filament loop. We assume that
the filament is the internal part of a twisted flux rope, which
has the shape of a half torus, with both ends staying anchored in
the chromosphere. The loop does not belong to a flat plane but
lies in the curved neutral surface (Fig.~\ref{Fig8}). In
accordance with the shape of the neutral surface, the fore leg of
the arch is more curved. The arch apex was slowly ascending during
several days before the eruption and reached a very high altitude
by December 2. We note that a quite similar schema has been
recently proposed by Alexander et al. (\cite{Alexander}) to
describe what he called a ``failed" filament eruption with a flare
and RHESSI hard X-ray emissions but no CME, so that we should
still be cautious with the association of an EP and a CME.  On the
other hand, this scenario agrees rather well with the recent model
of Chen et al. (\cite{Chen}), proposed to describe the CME-EP
phenomenon.

The dark feature visible before 09 UT in EUV images, which
coincides in space with the H$\alpha$ prominence, overlaps the
middle part and the far leg of the filament arch along the line of
sight. But the activation and the ascent of the prominence began
near its fore leg. The rising part becomes bright in EUV as is
observed in many eruptive prominences (Filippov \& Koutchmy
\cite{Filippov02}), while the far part of the arch is still at
rest. That is why we see rising bright features on the background
of the rather static dark formation. This frontal part of the
filament becomes clearly visible in the EIT images in emission at
09:48 UT and later. It is also clearly seen in the H$\alpha$
images of the eruptive prominence as a curved feature
(Fig.~\ref{Fig3}b-c).

The appearance of moving bright features within the body of an EP
is consistent with the mechanism of heating of the prominence
material proposed by Filippov \& Koutchmy (\cite{Filippov02}). The
source of energy in this mechanism is the magnetic energy of the
pre-eruptive magnetic configuration stored in the flux rope
electric current and its interaction with the surrounding fields
and plasma. The magnetic energy transforms into the macroscopic
kinetic energy of the filament motion, including its threads. Part
of this energy turns into heat through collisions inside the flux
tube after an increase of the turbulence inside. A rather complex
structure of the flux tube is necessary for this purpose. Some
segments of the tube should have an upward concave form to provide
the counter flow acceleration. We note that a typical twisting
flux rope easily meets this requirement and that such twist was
indeed observed in several instances (see e.g. Plunkett et al.
\cite{Plunkett}; Koutchmy et al. \cite{Koutchmy04}; Sterling \&
Moore \cite{Sterling}; Anderson et al. \cite{Anderson}). It is
also the requirement deduced from the latest theoretical analysis
of the flux rope model proposed for CMEs by Chen et al.
(\cite{Chen}).

\section{Conclusions}

We documented a large and fast coronal limb event, which occurred
on December 2, 2003, in the interval 9-14$^{\mbox{\small h}}$ UT,
in the vicinity of an old multi-polar region. Using observations
obtained on the ground and in space with two EUV telescopes, an
H$\alpha$ coronagraph and the LASCO white-light coronagraphs, we
analysed a spectacular dynamical phenomenon passing through
different stages: a quiescent filament initially, then a
prominence with EUV brightening at its periphery, its dynamical
and thermal activation, a coronal eruption with a small flare and
the launch of the eruptive prominence (EP) and finally the
formation of a classical three-part fast CME. This CME fits rather
well  within the framework of the so-called ``standard model",
(see e.g. Lin \cite{Lin}). Here we have added several features
that we believe could shed some light on the whole process,
emphasizing the mass transport processes related to the EP and its
coronal surroundings.

From what we saw, the eruption process apparently started with the
emergence of a new magnetic loop striking the prominence and
destabilizing it. This led to a rapid heating of the gas around
the prominence body, which spatially corresponds to a part of the
EUV filament channel. The brightness of the body of the prominence
in H$\alpha$ at first increased to a maximum value (at 09:35 UT)
possibly due to the growth of turbulence inside which produces a
large Doppler brightening effect and, accordingly, magnifies the
H$\alpha$ flux without the need to increase the densities. This
brightening even precedes the lifting of the cool prominence,
which seems to be an important new feature to be taken into
account in future work on CMEs. We note that this observation was
made possible by the use of a coronagraph having a broad band
filter and not a narrow band H$\alpha$ filter as is usually done
with a solar telescope to reject the parasitic light. After
09:41~UT the brightness decreased in H$\alpha$ and increased in
EUV, due to the enhanced ionization process and possibly a
dissipation of the mechanical energy from waves or counter-flow
motions. The prominence began to move outward at 09:48 UT, which
coincides with the X-ray flare peak and would precede the CME
extrapolated onset time.

A temporal analysis of the eruptive prominence motion and of the
dimming intensity light curves has shown that both the frontal
structure and the core of a CME were initiated simultaneously
within 5-10 min around the peak of an X-ray flare.

Evaluations based on photometric data coming from the W-L
coronagraphic coverage of the CME and from the photometric data
provided by the H$\alpha$  line-flux images of the prominence
before its launch show that the total mass of the CME almost
certainly surpasses the mass of the ejected prominence material
seen in H$\alpha$ before its eruption:
\begin{itemize}
\item   the mass of the CME core is 2-2.5 times greater than
the mass of the quiescent prominence, including the mass of an EUV
channel (although taking into account possible turbulence this
difference may vanish). This clearly shows that the mass of a fast
CME is mainly due to the surrounding low corona swept out during
the eruption, as confirmed by the large dimming effect we
measured. This part of the CME is not treated in this paper but
new ways already exist to consider it, e.g.  Delann\'{e}e et al.
(\cite{Del}).
\item   the dimming effect is observed over a large part of the
corona, on the disk and also outside the disk when the line of
sight integration reaches a 2-fold increased value right above the
limb with a corresponding increase of the dimming effect.
 The dimming is in phase with the onset of the CME. Approximately 5\%
or more of the corona is depleted, neglecting the depletion of the
mass of the corona at very low (TR) and/or very high coronal
temperatures. The corona seen in the EIT 195 \AA\ channel should
indeed correspond to the bulk of the mass because it corresponds
to the most probable temperature of the corona, so that our
estimate should not be far from the truth.
\end{itemize}

We also found that the eruption led to a spatial deflection and a
brightening of the neighboring South-West streamer, which may give
rise to an overestimate of the CME mass by  $\sim~20$\%. Further
out toward both polar regions the deflections of structures such
as  extended plumes are clearly visible as a back and forth
quasi-coherent transverse motion. This phenomenon would need
special study which is beyond the scope of this paper.

Finally our results suggest that greater attention should be paid
to regions around the EP, bringing added justification to the
theoretical studies trying to describe the behavior of the coronal
magnetic field leading to CMEs. However only the field inside
prominences can be easily measured with cool lines and the Hanle
effect; an improved precision H$\alpha$ coronagraph could
routinely be used to perform such measurements at high speed.

\begin{acknowledgements}
The authors are grateful to O.G. Den for coronal magnetic field
calculations, to J.-C. Vial for clarifications concerning the
Doppler brightening effect, to G. Stellmacher for in-depth
discussions on the determination of the prominence mass and the
interpretation of prominence observations, to C. Delann\'{e}e, G.
Lawrence for discussions on CMEs and to A. Zhukov for carefully
looking at the 1st draft of the paper. This work was supported in
part by the Russian Foundation for Basic Research (grants
06-02-16424 and 05-02-17415) and in part by the NATO CRG 940291.
SOHO is a space project of international cooperation between ESA
and NASA and we specially thank the teams supporting the EIT and
the LASCO observations for excellent work and providing their data
to the community. S.K. also benefited from CNES in collaborating
with the LASCO team in Marseille. The HACO ground-based coronal
H$\alpha$ observations are supported by the ``Observateurs
Associ\'{e}s" team sponsored by Fiducial LTD and by the
``Observatoire Midi-Pyr\'{e}n\'{e}es". We are grateful  to M.
Audejean and J. Guignard for collecting the sequence used in this
paper. We are also indebted to the international H$\alpha$ Network
for providing filtergrams to cover the chromospheric activity in
2003.
\end{acknowledgements}


\begin{thebibliography}{}

\bibitem [2006]{Alexander}
Alexander, D., Lin, R., \& Gilbert, H. R.  2006, \apj, 653, 719

\bibitem [2002]{Aulanier}
Aulanier, G.  \&  Schmieder, B. 2002, \aap, 386, 1106

\bibitem [2005]{Anzer}
Anzer, U. \& Heinzel, P. 2005, \apj, 622, 714

\bibitem [1948]{d'Azambuja}
d'Azambuja, M. \& d'Azambuja, L. 1948, Ann. Obs. Paris, Meudon, V.
6, Fasc. VII

\bibitem [2007]{Bao}
Bao, X. Zhang, H. Lin, J., \& Stenborg, G. A. 2007, \aap, 463, 321

\bibitem [1994]{Batchelor}
Batchelor, D. A. \& Schmahl, E. J. 1994, Solar Dynamic Phenomena
and Solar Wind Consequences, (Ed.) J.J. Hunt, ESA SP-373, 203

\bibitem [1998]{Boer}
de Boer, C. R., Stellmacher, G., \& Wiehr, E. 1998, \aap, 334, 280

\bibitem [1994]{Bothmer}
Bothmer, V. \& Schwenn, R. 1994, \ssr, 70, 215

\bibitem [2006]{Chen}
Chen, J., Marqu\'{e}, C. Vourlidas, A. Krall, J., \& Schuck, P. W.
2006, \apj, 649, 452

\bibitem [2004]{Chertok04}
Chertok, I. M., Slemzin, V. A., Kuzin, S. V., et al. 2004,
Astronomy Reports, 48, 407

\bibitem [2005]{Chertok05}
Chertok, I. M. \& Grechnev, V. V. 2005, \solphys, 229, 95

\bibitem [2004]{Cremades}
Cremades, H.  \& Bothmer, V. 2004, \aap, 422, 307

\bibitem [1995]{Delaboudiniere}
Delaboudiniere, J.-P., Artzner, G. E., Brunaud, J. et al.  1995,
\solphys, 162, 291

\bibitem [2007]{Del}
Delann\'{e}e, C., Hochedez, J-F., \& Aulanier, G. 2007, \aap, in
press.

\bibitem [2002]{Den}
Den, O.G., 2002, Astronomy Letters, 28, 345

\bibitem [2000]{Filippov00}
Filippov, B. P. \& Den, O. G. 2000, Astronomy Letters 26, 322

\bibitem [2001]{Filippov01}
Filippov, B. P. \& Den, O. G., 2001, \jgr, 106, 25177

\bibitem [2002]{Filippov02}
Filippov, B. \& Koutchmy, S. 2002, \solphys, 208, 283

\bibitem [2007]{Filippov07}
Filippov, B. \& Zagnetko, A.  2007, Journal of Atmospheric and
Solar-Terrestrial Physics, in press.

\bibitem [1991]{Forbes}
Forbes, T. G. \& Isenberg, P. A. 1991, \apj, 373, 294

\bibitem [2000]{Gilbert00}
Gilbert H. R., Holzer T. E., Burkepile J. T., \& Hundhausen A. J.
2000, \apj, 537, 503

\bibitem [2005]{Gilbert05}
Gilbert, H. R., Falco, L. E., Holzer, T. E., \& MacQueen, R. M.
2005, \apj, 618, 524

\bibitem [2006]{Gilbert06}
Gilbert, H. R., Falco, L. E., Holzer, T. E., \& MacQueen, R. M.
2006, \apj, 641, 606

\bibitem [1997]{Gontikakis}
Gontikakis, C. Vial, J-C., \& Gouttebroze, P. 1997, \aap, 325, 803

\bibitem [2003]{Gopalswamy03}
Gopalswamy, N., Shimojo, M., Lu, W., et al. 2003, \apj, 586, 562

\bibitem [2006]{Gopalswamy06}
Gopalswamy, N. 2006, JA\&A, 27, 243

\bibitem [1993]{Gouttebroze}
Gouttebroze, P., Heinzel, P., \& Vial, J.-C. 1993, \aaps,  99, 513

\bibitem [2006]{Grechnev}
Grechnev, V. V., Chertok, I. M., Slemzin, V. A., et al. 2005,
\jgr, 110, A09S07

\bibitem [2001]{Harra}
Harra, L. K. \& Sterling, A. C. 2001, \apjl, 561, 215

\bibitem [2003]{Harrison}
Harrison, R. A., Bryans, P., Simnett, G. M., \& Lyons, M. 2003,
\aap, 400, 1071

\bibitem [1987]{Heinzel87}
Heinzel, P. \& Rompolt, B.  1987, \solphys, 110, 171

\bibitem [2001]{Heinzel01}
Heinzel, P., Schmieder, B., \& Tziotziou, K. 2001, \apj, 561, L223

\bibitem [1999]{Heinzel99}
Heinzel, P., Mein, N., \& Mein, P. 1999, \aap, 346, 322

\bibitem [2003a]{Heinzel03a}
Heinzel, P., Anzer, U., Schmieder, B., \& Schwarts, P. 2003a, ESA
SP-535, 447

\bibitem [2003b]{Heinzel03b}
Heinzel, P., Anzer, U., \& Schmieder, B. 2003b, \solphys, 216, 159

\bibitem [1991]{Herant}
Herant, M., Pardo, F., Spiller, E., \& Golub, L. 1991, \apj, 376,
797

\bibitem [1997]{Hudson}
Hudson, H. S. \& Webb, D. F. 1997, Coronal Mass Ejections, ed. N.
Crooker, J. Joselyn, \& J. Feynman (Washington, DC: AGU
Geophysical Monographs 99), 27.

\bibitem [2004]{Jing}
Jing, J., Yurchyshyn, V. B., Yang, G., Xu, Y., \& Wang, H. 2004,
\apj, 614, 1054

\bibitem [1981]{Koutchmy81}
Koutchmy, S. \& Nikolsky, G. M. 1981, Sov. Astron. Lett. 7, 102

\bibitem [2004]{Koutchmy04}
Koutchmy, S., Baudin, F., Bocchialini, K., Daniel, J.-Y., et al.
2004,  \aap, 420, 709

\bibitem [1998]{Kucera}
Kucera, T. A., Andretta, V., \& Poland, A. I. 1998, \solphys, 183,
107

\bibitem [2004]{Lin}
Lin, J. 2004, \solphys, 222, 115

\bibitem [2003]{Low}
Low, B. C., Fong, B., \& Fan, Y. 2003, \apj, 594, 1060

\bibitem [1996]{Mein96}
Mein, N., Mein, P., Heinzel, P., et al. 1996, \aap, 309, 275

\bibitem [2001]{Mein01}
Mein, N. Schmieder, B. DeLuca, E. E., et al. 2001, \apj, 556, 438

\bibitem [1987]{Molodensky}
Molodensky, M. M. \& Filippov, B. P., 1987, \sovast, 31, 564

\bibitem [2000]{Penn}
Penn, M. J. 2000, Sol. Phys., 197, 313

\bibitem [2000]{Plunkett}
Plunkett, S. P., Vourlidas, A., Simberova, S., et al., 2000,
\solphys, 194, 371

\bibitem [2007]{Romeuf}
Romeuf, D., Meunier, N., Noens, J.-C., et al.  2007, \aap,  462,
731.

\bibitem [2003]{Schmieder03}
Schmieder, B., Tziotziou, K., \& Heinzel, P. 2003, \aap, 401, 361

\bibitem [2004]{Schmieder04}
Schmieder, B., Lin, Y., Heinzel, P.. \& Schwartz, P. 2004,
\solphys, 221, 297

\bibitem [2006]{Schwartz}
Schwartz, P,. Heinzel, P., Schmieder, B., \& Anzer, U. 2006, \aap,
459, 651

\bibitem [2005]{Slemzin05}
Slemzin, V. A., Kuzin, S. V., Zhitnik, I. A., et al. 1995, Solar
System Research, 39, 489

\bibitem [2006]{Slemzin06}
Slemzin, V. A., Grechnev, V. V., \& Kuzin, S. V. 2006, Solar
Activity and its Magnetic Origin, Proceedings of the 233 IAU
Symposium, (Eds.) V. Bothmer \& A. A. Hady, Cambridge: Cambridge
University Press, p. 361

\bibitem [1990]{Soru-Escaut}
Soru-Escaut, I. \& Mouradian, Z. 1990, \aap, 230, 474

\bibitem [2004]{Sterling}
Sterling, A. C. \& Moore, R. L. 2004, \apj, 599, 1418

\bibitem [1974]{Tandberg-Hanssen}
Tandberg-Hanssen, E. 1974, Solar Prominences, D. Reidel Publ. Co.,
Dordrecht, Holland

\bibitem [1978]{Van Tend}
Van Tend, W. \& Kuperus, M., 1978, Solar Phys. 59, 115

\bibitem [2007]{van Ballegooijen}
van Ballegooijen, A. A. \& Mackay, D. H. 2007, \apj, 659, 1713

\bibitem [2000]{Vourlidas}
Vourlidas, A., Subramanian, P., Dere, K. P., \& Howard, R. A.
2000, \apj, 534, 456

\bibitem [1984]{Wagner}
Wagner, W. J. 1984, \araa, 22, 267

\bibitem [2007]{Wiehr}
Wiehr, E., Stellmacher, G., \& Hirzberger, J. 2007, \solphys, 240,
25

\bibitem [2005]{Zagnetko}
Zagnetko, A. M., Filippov, B. P., \& Den, O. G. 2005, Astron.
Reports, 49, 425

\bibitem [1999]{Zarro}
Zarro D. M., Sterling, A. C., Thompson, B. J., Hudson, H. S., \&
Nitta, N. 1999, \apj, 520, L139

\bibitem [2002]{Zhitnik}
Zhitnik, I. A., Bougaenko, O. I., Delaboudiniere, J.-P., et al.
2002. ESA SP-506 "Solar Variability: from  Core to Outer
Frontiers". 2002, 915

\bibitem [2007]{Zhukov}
Zhukov, A. N. \& Auch\`{e}re, F. 2007, \aap, 427, 705




\end{thebibliography}
\end{document}